\documentclass[aps,pre,twocolumn,superscriptaddress,floatfix,showpacs]{revtex4}
\usepackage[dvips]{graphicx}
\usepackage{amssymb}

\begin{document}

\title{Ray-wave correspondence in the nonlinear description of
stadium-cavity lasers}

\newcommand{\affiliationATR}{\affiliation{Department of Nonlinear
Science, ATR Wave Engineering Laboratories, 2-2-2 Hikaridai, Seika-cho,
Soraku-gun, Kyoto, 619-0288, Japan}}

\newcommand{\affiliationYale}{\affiliation{Department of Applied
    Physics, Yale University, P.O. Box 208284, New Haven, Connecticut
    06520-8284}}

\author{Susumu Shinohara}
\affiliationATR

\author{Takahisa Harayama}
\affiliationATR

\author{Hakan E. T\"ureci}
\affiliationYale

\author{A. Douglas Stone}
\affiliationYale

\begin{abstract}
We show that the solution of fully nonlinear lasing equations for
stadium cavities exhibits a highly directional emission pattern. This
directionality can be well explained by a ray-dynamical model, where
the dominant ray-escape dynamics is governed by the unstable manifolds
of the unstable short periodic orbits for the stadium
cavity. Investigating the cold-cavity modes relevant for the lasing,
we found that all of the high-$Q$ modes have the emission
directionality corresponding to that of the ray-dynamical model.
\end{abstract}


\maketitle

\newcommand{\real}{{\mathbb R}}
\newcommand{\complex}{{\mathbb C}}
\newcommand{\vecr}{{\mbox{\boldmath $r$}}}
\newcommand{\vecn}{{\mbox{\boldmath $n$}}}
\newcommand{\vece}{{\mbox{\boldmath $e$}}}
\newcommand{\degree}{^{\circ}}

Establishing a correspondence between the ray/classical picture and
the wave/quantum picture has been a fundamental problem in the field
of wave/quantum chaos \cite{QC}.
One encounters this problem when trying to understand the emission
properties from two-dimensional microcavity lasers.
In such lasers, as a way to extract highly directional emission, it
has been proposed to deform the cavity shape smoothly from perfect
circularity \cite{Nockel94,Nockel96a,Nockel96b,Nockel97,Gmachl}.
The result is that rays start to exhibit a variety of dynamics from
integrable to strongly chaotic, which is tunable by the deformation.

The ray picture has been providing a simple and intuitive method to
explain experimental observations of emission directionality.
For instance, emission directionality has been associated with the
existence of a periodic orbit with a particular geometry
\cite{Gmachl,Rex}, drastic shape-dependence of emission directionality
has been successfully explained by the difference of phase space
structure \cite{Schwefel}, and the far-field intensity patterns have
been closely reproduced by ray-tracing simulations
\cite{Schwefel,Hentschel,Fukushima}.

Among various cavity shapes, the stadium is a simple geometry for
which ray dynamics has been rigorously proven to become strongly
chaotic \cite{Bunimovich}.
For almost all initial conditions, a ray trajectory explores the
entire phase space uniformly.
Even for such a strongly chaotic cavity, if one considers refractive
emission of light due to the dielectric nature of the cavity, the
emission pattern can become highly directional.
Namely, strongly chaotic dynamics and highly directional emission are
compatible, as was demonstrated by Schwefel et al. \cite{Schwefel},
who associated this property with escape dynamics dominated by flow in
phase space along the unstable manifolds of the unstable short
periodic orbits of a chaotic system.

In this paper, we report further evidence for the ability of a
ray-dynamical model to describe the lasing states of two-dimensional
microcavities.
Earlier work has focused on establishing a relationship between the
ray model and a few quasi-bound state solutions of the linear wave
equation, without pumping or gain.
Which modes to choose for comparison in this case has an intrinsic
arbitrariness, although plausibility arguments can be made based on
their $Q$ values.
Here we show that the solution of the full nonlinear lasing equations
for a stadium cavity, uniquely determined by the pumping conditions,
has highly directional emission in good agreement with the ray model.
This is one of the first pieces of evidence that the multi-mode
solutions of nonlinear wave equations can be understood in terms of
the classical limit of its linear counterpart.

\begin{figure}[b]
\includegraphics[width=72mm]{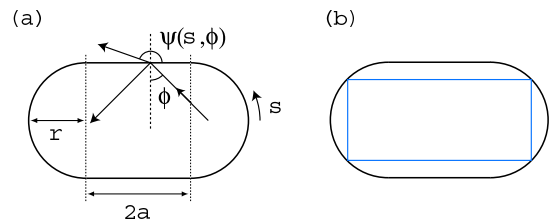}
\caption{(a) Geometry of the stadium cavity, (b) the rectangular
  unstable periodic orbit.}
\label{fig:stadium_cavity}
\end{figure}

\begin{figure}[t]
\includegraphics[width=80mm]{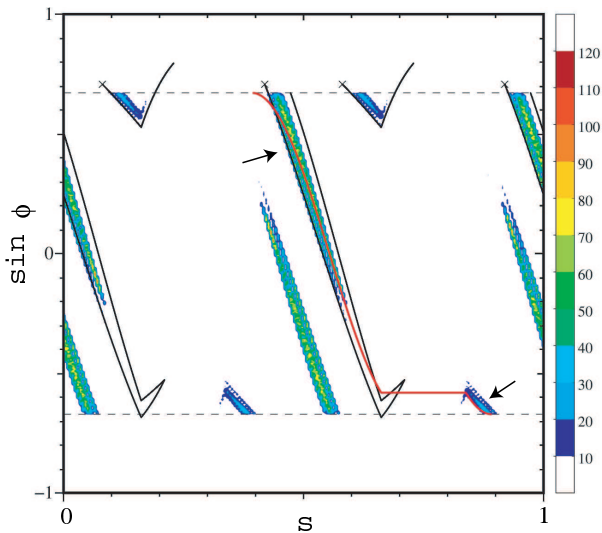}
\caption{Ray model simulation: intensity distribution of emitted rays,
  $I(s,\sin\phi)$. The black solid curves are the unstable manifolds
  of the rectangular unstable periodic orbit marked by X.  The red
  curve is a set of points giving the far-field emission at
  $\theta=210\degree$. The black dashed lines indicate the critical
  lines for total internal reflection defined by $\sin\phi=\pm
  1/n_{in}$. }
\label{fig:phase_space}
\end{figure}
\begin{figure}[t]
\includegraphics[width=79mm]{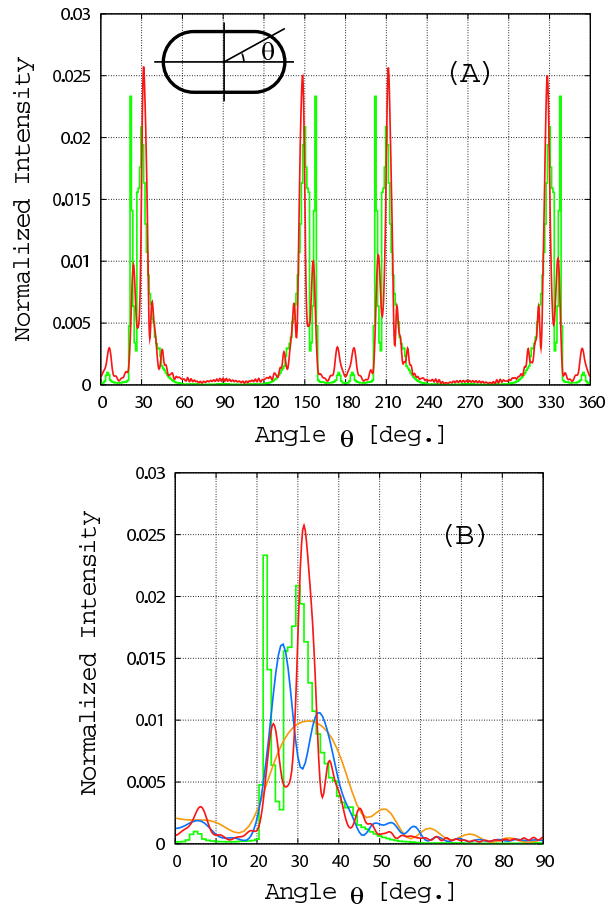}
\caption{The far-field intensity patterns for the stadium cavity with
  $n_{in}=1.49$. (A) The ray model (green) vs. the Schr\"odinger-Bloch
  model (red). (B) The ray model (green) vs. the Schr\"odinger-Bloch
  model (red: $r=67.77$, blue: $r=33.88$, orange: $r=16.94$).}
\label{fig:ffp.sb-vs-ray}
\end{figure}

Firstly, it is shown that a ray model for the stadium cavity exhibits
highly directional far-field emission.
In this paper, we fix the aspect ratio of the stadium to $a/r=6/7$,
where $r$ and $a$ are the radius of the semicircles and the
half-length of the straight segments, respectively
(Fig. \ref{fig:stadium_cavity} (a)).
In the ray model, the appearance of strong directionality depends
heavily on the value of the refractive index of the cavity $n_{in}$.
We set $n_{in}=1.49$, which corresponds to the index for polymer
cavities\cite{Schwefel}.

The ray model is constructed on the basis of Snell's and Fresnel's
laws \cite{Nockel96a, Nockel96b, Nockel97, Hentschel, Fukushima,
Schwefel, Lee04, Lee05, Ryu}.
Inside the cavity, the dynamics of a ray is viewed as the motion of a
point particle moving freely except for reflections at the cavity
boundary.
The ray dynamics can be reduced to a two-dimensional area-preserving
mapping by introducing the Birkhoff coordinates $(s,\sin\phi)$, where
$s$ is the arc-length along the cavity boundary and $\phi$ is the the
angle of incidence (Fig. \ref{fig:stadium_cavity}(a)).
Taking into account the dielectric nature of the cavity, we consider
the emission of rays to the outside of the cavity, which is done in
the following manner.
Each ray is initially assigned a certain amount of intensity.
This intensity decreases whenever a ray collides with the cavity
boundary, where the amount of the emitted-ray intensity is determined
by Fresnel's law, while the ray's emission angle is given by Snell's
law.

In the ray model simulations, we prepare the initial ensemble of rays
to be uniformly distributed in the phase space spanned by the Birkhoff
coordinates.
After some transient, the total intensity of rays inside the cavity
decreases exponentially as a function of time \cite{Ryu}.
In such a stationary regime, we measure the intensity distribution for
the emitted-rays $I(s,\sin\phi)$.

The qualitative explanation for strong emission directionality in such
a chaotic system was given by Schwefel et al \cite{Schwefel}; the
unstable short periodic orbits act like anisotropic "scattering
centers" in phase space, causing directional flow along their unstable
manifolds until the critical angle for escape is reached.
For shapes like the stadium there exists a "line of constant
far-field" corresponding to the set of values of the angle of
incidence $\phi$ and angular position on the boundary $s$ at which a
ray refracts in the same far-field angular direction.
The line of constant far-field closest to the unstable manifold then
predicts the dominant emission directionality.
Note that all of the relevant short orbits have closely nested
unstable manifolds so this gives a unique prediction \cite{Schwefel}.
The data in Fig 2 are in agreement with this picture.
In Fig. 2, $I(s,\sin\phi)$ is plotted overlaid with the unstable
manifolds of the rectangular unstable periodic orbit (Fig. 1(b))
located near the critical line for total internal reflection, i.e.,
$\sin\phi=1/n_{in}$.
For $n_{in}=1.49$, we see that high escape intensity regions consist
of narrow stripes corresponding closely to the unstable manifolds, and
that two stripes of high intensity (marked by arrows) are almost
parallel to the constant far-field curve for $\theta=210\degree$ (red
curve), where the angular coordinate $\theta$ is defined as the
counterclockwise angle from the $x$-axis.
By the four-fold symmetry of the stadium each of the other stripes is
parallel to one of the curves for $\theta=30\degree, 150\degree$ and
$330\degree$, giving a
four-fold symmetric directional emission pattern peaked around these
angular directions.
We plot the far-field intensity pattern ${\cal F}(\theta)$ in
Fig. \ref{fig:ffp.sb-vs-ray}, which can be calculated from the
intensity distribution $I(s,\sin\phi)$ via
%
%\begin{equation}
$
{\cal F}(\theta)=\int\int ds\,d(\sin\phi)
~I(s,\sin\phi)\,\delta(\psi(s,\phi)-\theta),
$
%\end{equation}
%
where $\psi(s,\phi)$ is the emission angle (measured from the
$x$-axis) at $s$ for a ray with the angle of incidence $\phi$
\cite{Lee05}.
This is the result of the ray model.

Now, we investigate whether one can find the same emission
directionality for the lasing states for this cavity.
We describe the light field by the Maxwell equations, and assume that
the active medium consists of two-level atoms obeying the Bloch
equations.
In the description using the Maxwell and Bloch equations, the field
variables oscillate rapidly at a frequency close to the transition
frequency $\omega_0$ of the two-level atoms.
To perform the long-term time-evolutions necessary to obtain
stationary lasing solutions, we employ the Schr\"odinger-Bloch (SB)
model, which describes the time-evolution of the slowly varying
envelopes of the field variables
\cite{Harayama03a,Harayama03b,Sunada,Harayama05}.
The SB model is given by
\begin{eqnarray}
\frac{\partial E}{\partial t} &=& \frac{i}{2}
\left(\nabla^2+\frac{n^2}{n_{in}^2}\right)E-\alpha
E +\mu\rho,
\label{eq:sb1}\\
\frac{\partial \rho}{\partial t} &=&
-\gamma_{\perp}\rho
+\kappa W E,
\label{eq:sb2}\\
\frac{\partial W}{\partial t} &=&
-\gamma_{\parallel}(W-W_{\infty})
-2\kappa(E \rho^* + E^* \rho),
\label{eq:sb3}
\end{eqnarray}
where $E(x,y,t)$ and $\rho(x,y,t)$ are the slowly varying envelope of
the TM electric field and that of the polarization field,
respectively, and $W(x,y,t)$ is the population inversion component.
The refractive index $n(x,y)$ is $n_{in} (=1.49)$ inside the cavity
and $1.0$ outside it, and the linear absorption coefficient
$\alpha(x,y)$ is $\alpha_L (=const)$ inside the cavity and zero
outside it.
Space and time are made dimensionless by the scale transformations
$(n_{in}\omega_0 x/c,n_{in}\omega_0 y/c)\to(x,y)$ and $\omega_0 t \to
t$.
$\gamma_{\perp}$ and $\gamma_{\parallel}$ are phenomenological
relaxation rates, $\kappa$ and $\mu$ are the coupling strength
between the light field and the active medium, and $W_{\infty}$
represents the pumping strength.

Unless otherwise mentioned, the cavity size is $r=67.77$, for which
the perimeter length of the stadium becomes about $105$ times as large
as the light wavelength inside the cavity, i.e., $\lambda_{in}=2\pi$.
The other parameter values are set as follows: $W_{\infty}=0.01$,
$\gamma_{\perp}=10^{-2}$, $\gamma_{\parallel}=10^{-5}$,
$\alpha_L=10^{-3}$, $\kappa=0.5$, $\mu=\pi/n_{in}^2$.
For the above choice of the parameter values, around $100$ cavity
modes have positive linear gain.
For such a condition, there occur complicated interactions between the
modes, such as mode-pulling and mode-pushing, that generally yield a
multi-mode lasing solution \cite{Harayama03b, Sunada}.
We are interested in the far-field patterns arising from these
multi-mode lasing solutions; we now formulate a convenient method for
finding this quantity on the basis of the field data just outside the
cavity.

Using the Wiener-Khinchin theorem, one can write the time-averaged
light intensity as
\begin{equation}
\overline{{\cal I}}(\vecr,t)=\int d\omega\,
\lim_{T\to\infty}\frac{2\pi}{T}
\left|
E_T(\vecr,\omega)
\right|^2,
\end{equation}
where $E_T(\vecr,\omega)=\frac{1}{2\pi}\int_{-T/2}^{T/2} E(\vecr,t)
e^{i\omega t}$.
In the far-field regime $(r\gg 1)$, we obtain
$|E_T(\vecr,\omega)|^2 = |f_T(\theta,\omega)|^2/r$,
%
%\begin{equation}
%|E_T(\vecr,\omega)|^2 = \frac{|f_T(\theta,\omega)|^2}{r},
%\end{equation}
%
where the amplitude $f_T(\theta,\omega)$ is given by
\begin{equation}
f_T(\theta, \omega)=\frac{1+i}{4\sqrt{\pi k}}\oint_{\cal C}\,ds\,
e^{-i k (\vece\cdot\vecr_s)}
~\vecn_s\cdot\left(i k \vece +\nabla \right)
E_T(\vecr_s,\omega),
\label{eq:a3}
\end{equation}
with $k=\sqrt{1/n_{in}^2+2\omega}$. 
The integration is performed along a closed curve ${\cal C}$
encircling the cavity, $\vecr_s$ denotes a point on ${\cal C}$,
$\vecn_s$ is an unit vector normal to the curve ${\cal C}$ at
$\vecr_s$, and $\vece=\vecr/r=(\cos\theta, \sin\theta)$.
We define the time-averaged far-field pattern $\overline{{\cal
F}}(\theta)$ as the angle-dependent part of $\overline{{\cal
I}}(\vecr,t)$, i.e.,
\begin{equation}
\overline{{\cal F}}(\theta)=\int d\omega\,
\lim_{T\to\infty}\frac{2\pi}{T}|f_T(\theta,\omega)|^2.
\end{equation}

In Fig. \ref{fig:ffp.sb-vs-ray}(A), we plot $\overline{{\cal
F}}(\theta)$ for the stationary lasing solution of the SB model.
One can see a strikingly good agreement between the result from the SB
model and that from the ray model.
It is remarkable that the far-field pattern of the SB model reproduces
not only the highest peaks, but also the tiny ones at $\theta\approx
5\degree, 175\degree, 185\degree$ and $355\degree$.
The magnification of the first quadrant is shown in
Fig. \ref{fig:ffp.sb-vs-ray}(B), where the results for the cavity
sizes $r=33.88$ and $16.94$ are also presented.
Since the increase of the $r$-value results in the decrease of the
wavelength, one can see that the larger the $r$-value, the shorter the
spatial oscillation period becomes.
Nevertheless, if we average out the oscillations, all these far-field
patterns show a similar trend with a peak at around
$\theta=30\degree$.
This invariance of the peak position with respect to the change of the
cavity size convinces us that our system with $r=67.77$ is well inside
the semiclassical regime.
\begin{figure}
\includegraphics[width=50mm]{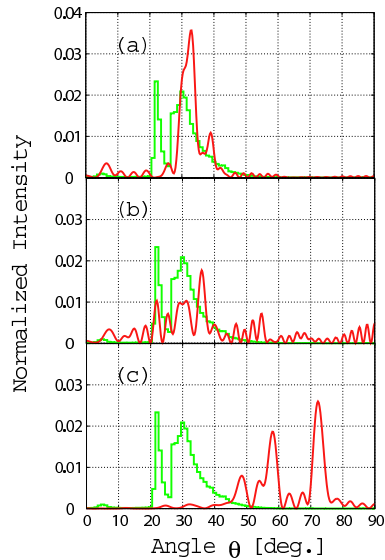}
\caption{The far-field patterns for the cold-cavity modes with
  even-even parity (solid line) and for the ray model (dashed
  line). (a) $\xi=-0.007054-0.002507\,i$, (b)
  $\xi=0.009670-0.005197\,i$, (c) $\xi=-0.006299-0.010132\,i$.
  Because of the pattern's symmetry, only the data in the first
  quadrant are shown.  }
\label{fig:ffp.cavity_modes}
\end{figure}

We note that besides the present study for $a/r=6/7$, we performed
numerical simulations of the SB model for various $a/r$-values ranging
from $0.13$ to $1.4$ while the other parameters are set to be the same
values as those used in this paper. For relatively large $a/r$-values,
we could confirm a good ray-wave correspondence. A detailed study on
the condition for the good ray-wave correspondence will be reported
elsewhere.

One natural approach to explain the appearance of the emission
directionality in the SB model is based on the study of the
cold-cavity modes.
Analyzing the power spectrum for the electric field of the lasing
solution for $r=67.77$, we confirmed that it consists of multiple
lasing modes, with the major contribution being from six lasing modes,
and these all exhibit strong far-field emission at $\theta\approx
30^{\circ}, 150^{\circ}, 210^{\circ}$ and $330^{\circ}$.
From this result, we can infer that cold-cavity modes with the above
directionality are preferentially excited.
To investigate the origin of this mode selection, we study below the
cold-cavity modes, focusing on the dependence of the emission
directionality upon the $Q$-value, which is one of the important
factors for the mode selection.

In the SB model, a cold-cavity mode, $E(x,y,t)=e^{-i\xi
 t}\psi(x,y)\,(\xi\in{\mathbb C})$, is a solution of
 Eq. (\ref{eq:sb1}) with $\alpha=\mu=0$.
Namely, $\psi(x,y)$ satisfies
%
%\begin{equation}
$
\left(
\nabla^2+\frac{n^2}{n_{in}^2}+2\xi
\right) \psi(x,y)=0.
$
%\end{equation}
%
%%We restrict our attention to even-even modes, i.e., those satisfying
%%$\phi(-x,-y)=\phi(x,y)$.
%%
%%In the gain band of our simulation, i.e., $|\mbox{Re}\,\xi|\leq 0.01$,
%%we find 23 even-even modes numerically.
%
We plot in Fig. \ref{fig:ffp.cavity_modes}(a) the far-field pattern of
a cold-cavity mode having strong far-field emission at $\theta\approx
30\degree$.
In the gain band of our simulation, i.e., $|\mbox{Re}\,\xi|\leq 0.01$,
we find 98 cold-cavity modes numerically.
As Figs. \ref{fig:ffp.cavity_modes}(b) and (c) show, there also exist
cold-cavity modes whose far-field patterns are less or even not
similar with that of the ray model.
Such modes, however, turn out to have lower $Q$ values as we see
below.

\begin{figure}
\includegraphics[width=80mm]{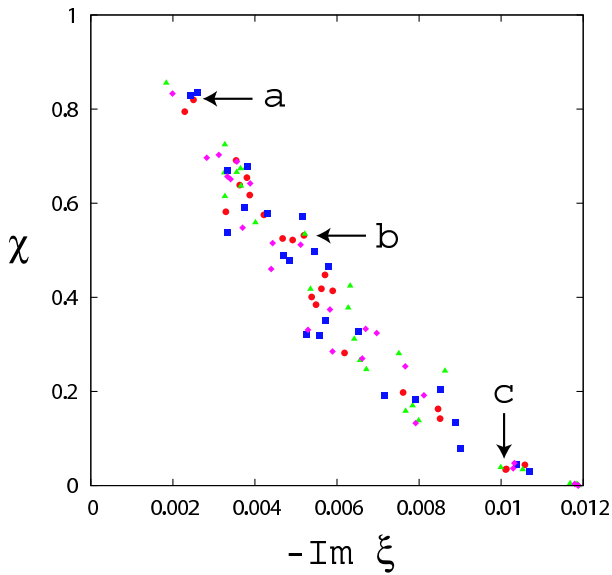}
\caption{The dependence of $\chi$ (far-field emission strength around
  $\theta=30\degree$) on the loss rate $-\mbox{Im}\,\xi$ for even-even
  (circle), even-odd (triangle), odd-even (square) and odd-odd
  (diamond) modes. The modes corresponding to the far-field patterns
  in Fig. \ref{fig:ffp.cavity_modes} are marked with arrows.}
%
%%\caption{The dependence of $\chi$ (far-field emission strength around
%%  $\theta=30\degree$) on the loss rate $-\mbox{Im}\,\xi$. The data is
%%  for even-even cavity modes. The modes presented in
%%  Fig. \ref{fig:ffp.cavity_modes} are marked with arrows.}
\label{fig:chi-values}
\end{figure}

To quantify the far-field emission strength around $\theta=30\degree$,
we compute the quantity
%
%\begin{equation}
$
\chi = \int_{20}^{40}
{\cal I}(\theta)\, d\theta \Bigm/ \int_{0}^{90} {\cal
I}(\theta)\,d\theta,
$
%\label{eq:chi}
%\end{equation}
%
where ${\cal I}(\theta)$ is the far-field pattern of the cold-cavity
mode.
Plotting $\chi$-values as a function of the loss rates
$-\mbox{Im}\,\xi$ as shown in Fig. \ref{fig:chi-values}, we find a
clear tendency that the lower the loss rate, the stronger the
far-field emission at $\theta\approx 30\degree$.
In other words, at least in this case the directional modes based on
the unstable manifolds also correspond to the high-$Q$ modes and are
thus preferentially selected for lasing, while in general we expect
that other factors besides the $Q$ value will come into the mode
selection, such as mode volume and spatial hole-burning effects.

We note that recently Lebental et al. have succeeded in experimentally
observing highly directional emission for a polymer stadium cavity
with the aspect ratio $a/r=0.8$\cite{Lebental}.
In the experiment, strong far-field emission has been observed at
$\theta\approx 30\degree, 150\degree, 210\degree$ and $330\degree$,
which agrees with our numerical results shown in
Fig. \ref{fig:ffp.sb-vs-ray}, although the aspect ratio for the
numerical simulation ($a/r=6/7$) is slightly different from $a/r=0.8$.
Carrying out numerical simulations of the SB model also for $a/r=0.8$,
we checked that this slight difference of the cavity geometry does not
cause a significant change in the peak position of the far-field
pattern; the peaks are shifted by only about $2\degree$.
%
%Carrying out numerical simulations of the SB model also for $a/r=0.8$,
%we checked that by this slight difference of the $a/r$-value the
%far-field angles for strong emission are shifted by around $2\degree$
%from those for $a/r\approx 6/7$.
%
A conspicuous difference between the numerical and experimental
far-field data is that one of the peaks in the experimental data
(Fig. 4 in Ref. \cite{Lebental}) has a three times larger intensity
than the other three peaks.
This is however due to the lifetime effect of the laser dye used in
the experiment.

In summary, we demonstrated via numerical simulations of the nonlinear
lasing equations that the stadium-cavity laser exhibits a highly
directional emission pattern in good agreement with the ray model,
which predicts emission directionality based on the geometry of the
unstable manifolds of short periodic orbits.
Furthermore we confirmed that for the stadium in this parameter range
all of the high-$Q$ modes exhibit this high emission directionality.
Further analysis is needed to elucidate to what extent this property
of the high-$Q$ modes holds when one changes the refractive index
value, cavity size, cavity shape and so on.

We acknowledge helpful discussions with Harald G.L. Schwefel and
Satoshi Sunada.
The work at ATR was supported in part by the National Institute of
Information and Communication Technology of Japan.
The Yale portion of this work was supported by the National Science
Foundation under Grant No. 0408638.


\begin{thebibliography}{99}
%
\bibitem{QC} M.C. Gutzwiller, {\it Chaos in Classical and Quantum
  Mechanics} (Springer, Berlin, 1990); H.J. Stockmann, {\it Quantum
  Chaos: An Introduction} (Cambridge University Press, Cambridge,
  England, 1999).
%
\bibitem{Nockel94} J.U. N\"ockel, A.D. Stone, and R.K. Chang,
  Opt. Lett. {\bf 19}, 1693 (1994)
%
\bibitem{Nockel96a} J.U. N\"ockel, A.D. Stone, G. Chen, H.L. Grossman,
  and R.K. Chang, Opt. Lett. {\bf 21}, 1609 (1996).
%
\bibitem{Nockel96b} J.U. N\"ockel and A.D. Stone, in {\it Optical
  Processes in Microcavities}, R.K. Chang and A.J. Campillo,
  eds. (World Scientific, Singapore, 1996).
%
\bibitem {Nockel97} J.U. N\"ockel and A.D. Stone, Nature {\bf 385}, 45 (1997).
%
\bibitem{Gmachl} C. Gmachl, F. Capasso, E.E. Narimanov, J.U. N\"ockel,
  A.D. Stone, J. Faist, D.L. Sivco, and A.Y. Cho, Science {\bf 98},
  1556 (1998).
%
\bibitem{Rex} N.B. Rex, H.E. Tureci, H.G.L. Schwefel, R.K. Chang, and
  A.D. Stone, Phys. Rev. Lett. {\bf 88}, 094102 (2002).
%
\bibitem{Schwefel} H.G.L. Schwefel, N.B. Rex, H.E. Tureci, R.K. Chang,
  A.D. Stone, T.B.-Messaoud, and J. Zyss, J. Opt. Soc. Am. B {\bf 21},
  923 (2004).
%
\bibitem{Hentschel} M. Hentschel and M. Vojta, Opt. Lett. {\bf 26},
  1764 (2001).
%
\bibitem{Fukushima} T. Fukushima and T. Harayama, IEEE J. Quantum
  Electron. {\bf 10}, 1039 (2004).
%
\bibitem{Bunimovich} L.A. Bunimovich, Commun. Math. Phys. {\bf 65},
  295 (1977).
%
%\bibitem{Bunimovich} L.A. Bunimovich, Funkt. Anal. Appl. {\bf 8}, 254
%(1974).
%
%\bibitem{Benettin} G. Benettin and J.-M. Strelcyn, Phys. Rev. A {\bf
%17}, 773 (1978); Ch. Dellago and H.A. Posch, Phys. Rev. E {\bf 52},
%  2401 (1995).
%
\bibitem{Lee04} S.-Y. Lee, S. Rim, J.-W. Ryu, T.-Y. Kwon, M. Choi, and
  C.-M. Kim, Phys. Rev. Lett. {\bf 93}, 164102 (2004).
%
\bibitem{Lee05} S.-Y. Lee, J.-W. Ryu, T.-Y. Kwon, S. Rim, and
  C.-M. Kim, Phys. Rev. A {\bf 72}, 061801(R) (2005).
%
\bibitem{Ryu} J.-W. Ryu, S.-Y. Lee, C.-M. Kim, and Y.-J. Park,
  Phys. Rev. E {\bf 73}, 036207 (2006).
%
%\bibitem{Hecht} E. Hecht, {\it Optics} (Addison-Wesley, 1987).
%
\bibitem{Harayama03a} T. Harayama, P. Davis, and K.S. Ikeda,
  Phys. Rev. Lett. {\bf 90}, 063901 (2003).
%
\bibitem{Harayama03b} T. Harayama, T. Fukushima, S. Sunada, and
  K.S. Ikeda, Phys. Rev. Lett. {\bf 91}, 073903 (2003).
%
\bibitem{Sunada} S. Sunada, T. Harayama, and K.S. Ikeda, Phys. Rev. E {\bf
  71}, 046209 (2005).
%
\bibitem{Harayama05} T. Harayama, S. Sunada, and K.S. Ikeda,
  Phys. Rev. A {\bf 72}, 013803 (2005).
%
\bibitem{Lebental} M. Lebental, J.S. Lauret, R. Hierle, and J. Zyss,
  Appl. Phys. Lett. {\bf 88}, 031108 (2006).
%
\end{thebibliography}
\end{document}